# Atmospheric Circulation Influence During Winter on Change of Air Pressure, Temperature and Spectral Transparency at Yakutsk Array

S. P. Knurenko[1, a], I. S. Petrov[1, b]

[1]Yu. G. Shafer Institute of Cosmophysical Research and Aeronomy SB RAS, 31 Lenin ave. Yakutsk, Russia.

## ABSTRACT

The paper presents long-term observations of the atmosphere in Yakutsk region. Analysis of the data for 40 year period indicates a gradual strengthening of cyclonic activity in the region and hence the increase of the average winter temperature, increase variations of the rest atmosphere, which greatly softens the continental climate of Central Yakutia.

## 1. INTRODUCTION

Historically, the study of the processes occurring in the atmosphere took place over the decades. Now, it is difficult to find science that is not using the information about the atmosphere and its parameters. Knowledge of geophysical parameters of the atmosphere is important in the study of ultrahigh energy cosmic rays. For example, when evaluating the total number of electrons and muons in extensive air shower $N_s$, $N_\mu$ and primary energy $E_0$ shower, which is determined by the total flux of EAS Cherenkov light. Weather variation is more important in conditions of extreme continental climate (continental subarctic climate), when parameters of atmosphere is not standard and changes dramatically over time. In Yakutsk transition between winter and summer and vice versa is especially important, equalization of temperature by height and air density is in good correlation with standard model. Under these conditions, processes of air showers producing either slowed or accelerated depending on the height and radiation from air showers in optical wavelength range 400 – 600 nm undergoes multiple scattering and more rapidly absorbed than in the standard atmosphere. Such a distribution of temperature and density of the atmosphere in height leads to an increase of the total flux of air showers Cherenkov light that may affect the evaluation of the shower energy determined by Cherenkov radiation [1]. All this is necessary to take into account for cosmic ray experimental data analysis.

[a] s.p.knurenko@ikfia.sbras.ru  [b]igor.petrov@ikfia.sbras.ru

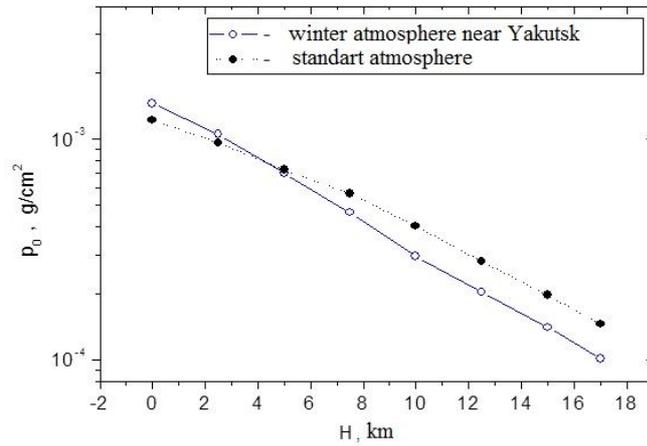

Fig. 1. Dependence of air density from the height for different models of the atmosphere.

## 2. YAKUTSK EXTENSIVE AIR SHOWERS ARRAY

Fig. 2 shows location of the detectors, Fig. 3 shows flowchart of Yakutsk array LAN.

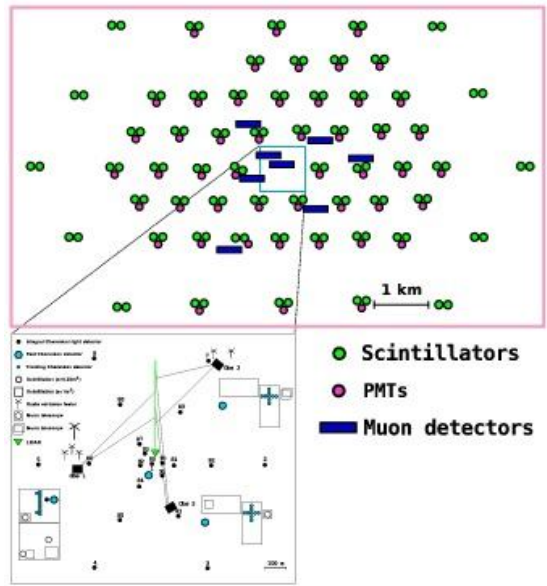

Fig. 2. Location of scintillation and Cherenkov detectors at Yakutsk array.

Yakutsk array consist several arrays [2]. All of them are designed to register electron, muon and Cherenkov components of air showers. Integral and differential Cherenkov detectors are used for measurements in the optical wavelength. Cherenkov detectors uses PMT with a large photocathode sensitive in the wavelength 300-600 nm, with a peak sensitivity at $\lambda = 430$ nm.

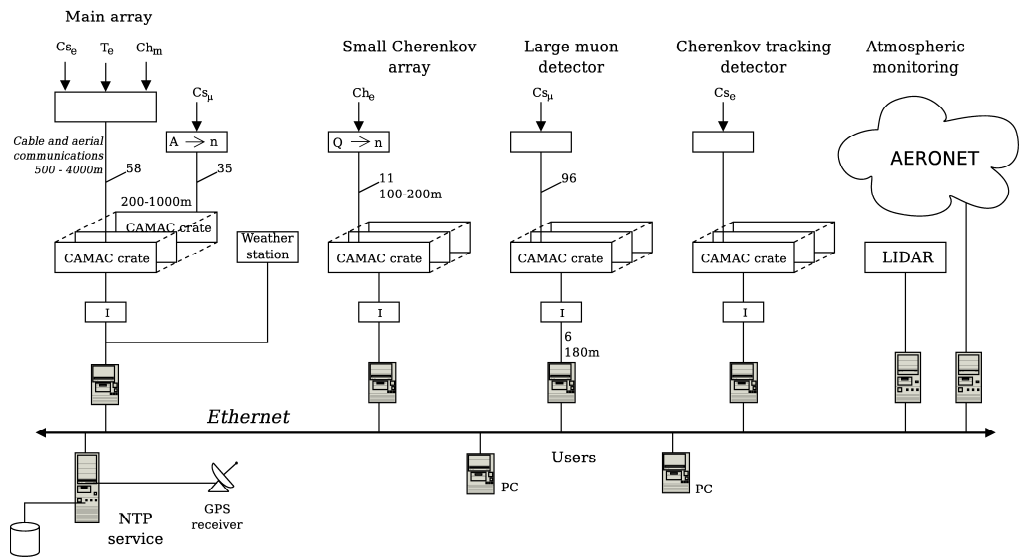

Fig. 3. Local area network of the Yakutsk array

The Small Cherenkov Array uses triangle grid system to register events of air showers [3]. The system has a good selection of showers frequency resolution up to 2000 Hz. A single measurement channel (ADC) allows measurements of the amplitude in the linear regime from 1mV to 10 V. As shown synchronous measurements of laser pulse by LIDAR telescope and Cherenkov detectors such equipment is enough to detect single pulses from a laser. Subsequently, differential Cherenkov detector was chosen as the target of the scattered laser radiation for intercalibration and assessment of height distribution haze and fog at temperatures below - 40 ° C.

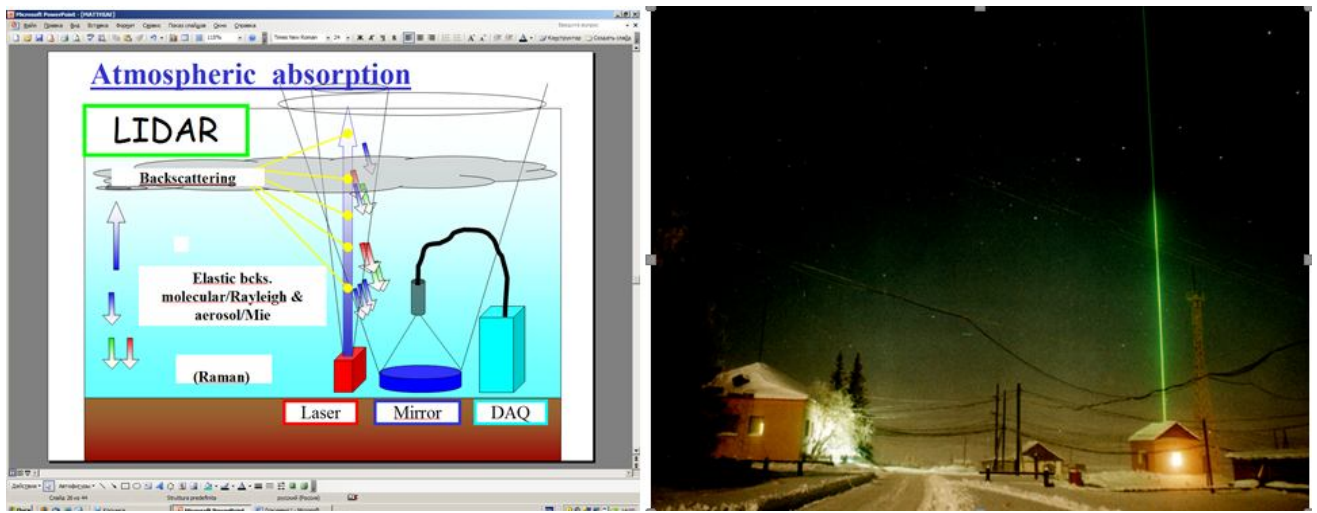

Fig. 4. The principle of operation of the optical array (a).   LIDAR laser beam (b)

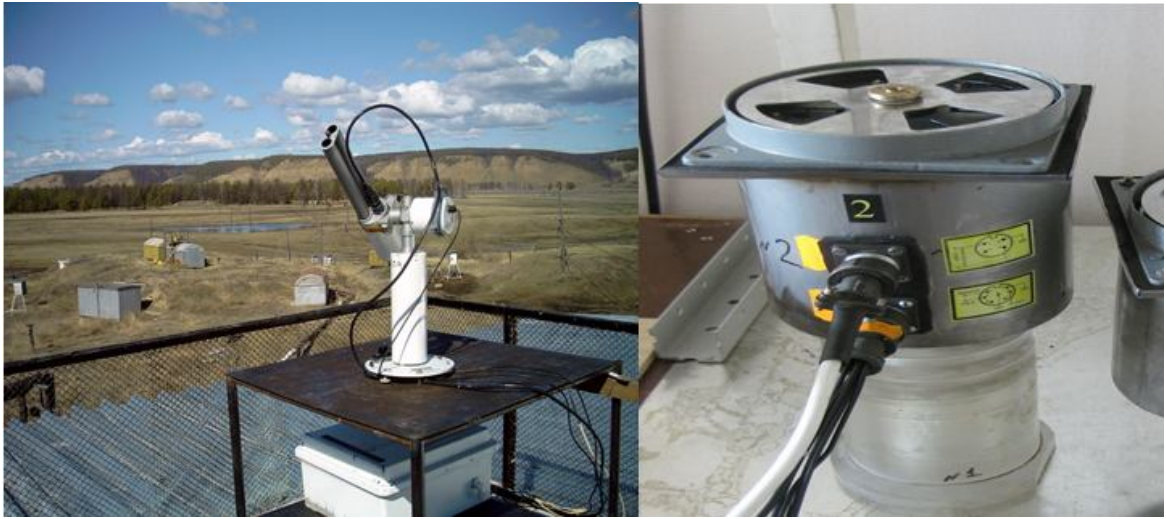
Fig.5. Photometer CE 318 at the top of station. Fluxmeter (b)

In addition to this, since 2003 a set up station of near atmosphere environmental monitoring at the site of the Yakutsk air shower array is operating. The station consists: a) Small Cherenkov array of extensive air showers that uses EAS event selection as the trigger b) LIDAR – performs transparency of the atmosphere monitoring (Fig. 4) c) Photometer CE 318 registers aerosol composition in the atmosphere (Fig. 5). The measuring system also includes electric field sensor (Fig. 5), antennas for measuring E and H components of the electric field, the direction finder of thunderstorm activity and scintillator-based thin plates for measurement of gamma and beta radiation. Permanently organized record of atmospheric noise field signals in the optical wavelength range at night and background variations of charged particles of different energies with scintillation detectors placed on the surface and background.

## 3. EXPERIMENTAL DATA

Figure 5a, 5b and 6a shows data for atmospheric monitoring observation periods ranging from 1970 to 2000 values of monitored parameters are grouped for each month and then averaged annualized. This analysis applies only to the winter months. For Yakutia these months are with negative temperatures from October to April. Clustered statistical data series have temperature variation in average annual values. However, in decadal intervals and interannual noticeable positive gradient, which is directed towards the increase of the average winter temperature. To clarify the effect we used broader observation period up to 2012 (see Fig. 7) [1]. Fig. 8 shows that there is a systematic increase ~0,8±0,2° C / decade of average winter temperature.

Atmospheric pressure during these periods is in antiphase to the temperature and, as seen from Fig. 6 also subject of considerable fluctuations. This is partly due to cyclonic activity in the atmosphere, changing modes cyclone - anticyclone and the onset of stable anticyclones coming from northern directions.

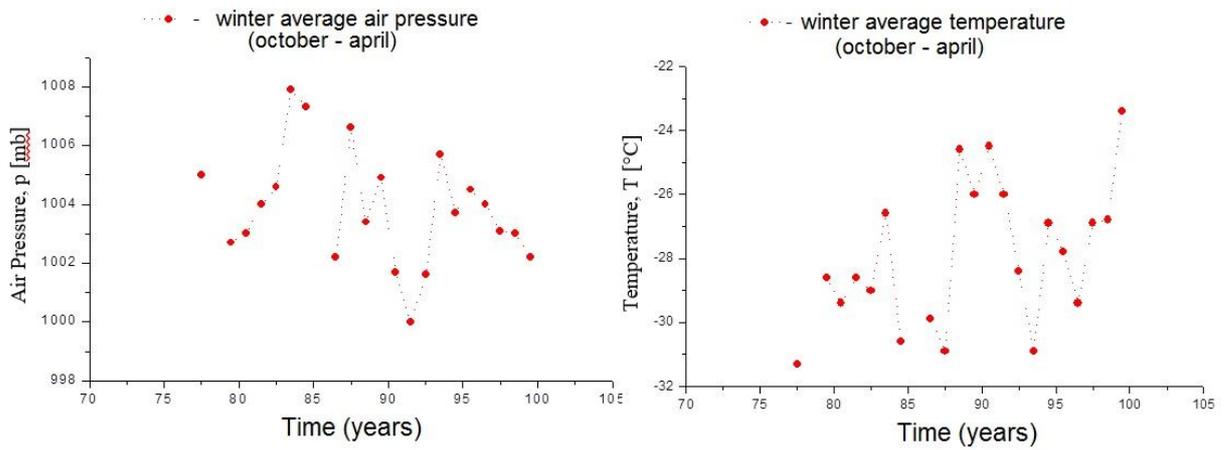

Fig. 6. Distribution of average winter temperature near Yakutsk array (a).
Changes of atmospheric pressure by years of observation (b).

From Fig. 7, it is clear that over the past 10 years number of cyclones from eastern and southern directions has increased. Quantitative evaluation of the integral coefficient of transparency for different optical states of the atmosphere was obtained by us in [4, 5] points directly to it. Here we used an adaptive method for solving inverse problems. As a result, we found a dependence of the transmittance of the atmosphere and height of the source (flash of Cherenkov radiation) in the atmosphere under different weather conditions.

From the analysis of many years data seen substantial fluctuations in the transmission coefficient of the atmosphere, which can significantly affect estimation of EAS parameters and to consider in the selection process and individual showers for analysis (Fig. 7). Therefore, to avoid additional errors in the analysis we selected EAS events when weather conditions were good and excellent, i.e. at $P_\lambda > 0,58$ [4].

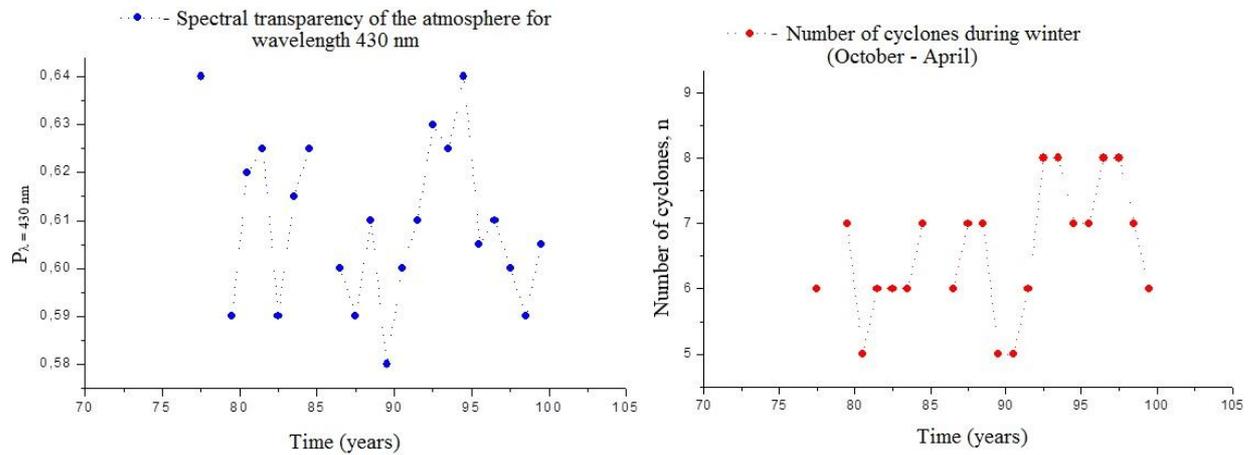

Fig. 7. Average transmission coefficient of the atmosphere, measured during winter at Yakutsk Array (a). Frequency of cyclones during winter near Yakutsk array area (b).

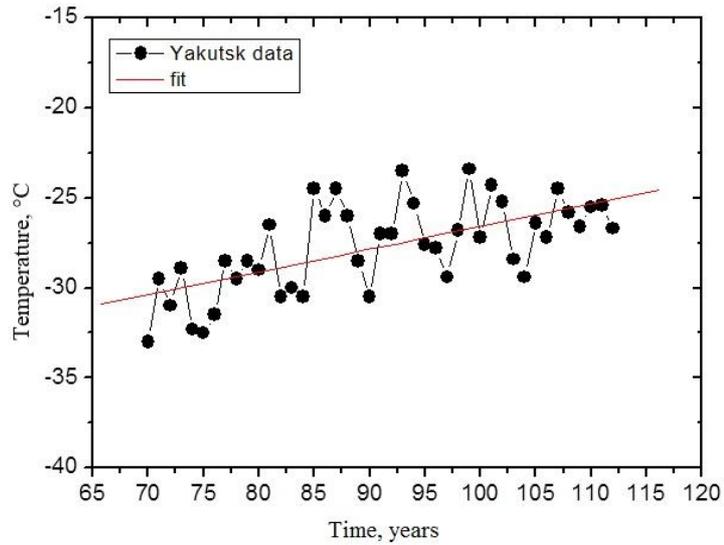

Fig.8. Temperature observation data over 40 years during winter near Yakutsk array area.

## 4. SOLAR AND GALACTIC COSMIC RAYS.

From the data of the atmosphere characteristics, we compared them with solar activity and galactic cosmic rays flux. Fig 8a shows data on solar flare activity (ICSU World Data System) and at Fig. 9 shows the intensity of GCR from 1970 to 2001. Data of variation of GCR obtained from Yakutsk observation [7].

The period covers three solar cycles from 1970 to the present time. During this same time, analyzes and observational material.

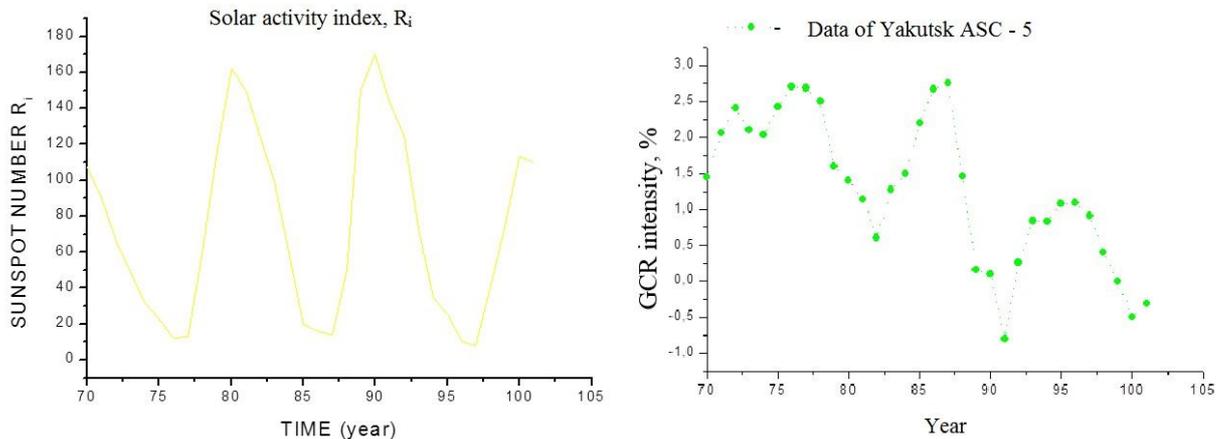

Fig. 9. Plot of average number of solar flares (a). Variations of galactic cosmic rays (b).

Table 1 shows the mean values of T, p, $P_\lambda$ and $n_i$ for winter period (October - April), where T - surface air temperature, p - pressure. $P_\lambda$ - spectral transparency of the lower atmosphere. The data are grouped by period of 10 years. Separately given line for the length of time from 1995 to 2001, where there is an abrupt change all of the parameters in Table 1.

Table 1. Average for the 10 year cycle characteristics of the atmosphere in region of Erkeeni valley.

| Period, year | ⟨T⟩, ° | ⟨p⟩, mb | ⟨P$\lambda$=430⟩ | $n_i$ |
|---|---|---|---|---|
| 1970 - 1980 | -31,0 | 1004,8 | 0,625 | 5 |
| 1980 - 1990 | -28,3 | 1004,0 | 0,610 | 6 |
| 1990 - 2001 | -27,5 | 1003,3 | 0,615 | 7 |
| 1995 - 2001 | -26,6 | 1003,0 | 0,60 | 8 |

From Table 1, it follows that there is a systematic increase of ~ 0,8 °C of average winter temperature and the pressure is also reducing. It seems like there is a redistribution of air density in lower troposphere as it is "warming up" the atmosphere, which may lead to deterioration and again an average transmittance of optical radiation of the atmosphere. In addition, table 1 shows that for last 10 years the number of cyclones is slightly increased. This factor further degrades the propagation of the optical wave in the lower atmosphere, as shown in Table 1.

Table 2 gives a qualitative picture of the correlation of solar activity index $R_i$ and the intensity of galactic cosmic rays $I_{gkl}$ with geophysical and optical parameters of the atmosphere.

Table 2. Correlation parameters of atmosphere with Ri and Igcr.

| Physical phenomenon | ⟨T⟩, ° | ⟨p⟩, mb | ⟨P$\lambda$=430⟩ | $n_i$ |
|---|---|---|---|---|
| Solar activity | C | 0 | AnC | AnC |
| Intensity of GCR | AnC | C | C | C |

Here, K, AnC and 0 indicates a correlation, anticorrelation and no correlation.

The results shown in Table. 2 were obtained by comparing the data shown in Fig. 6 and Figure 9.

The data (see Table 2) show that after the maximum of solar activity in 1999 and 2001 ÷ surface layers of air temperature decreased by 2.4 degrees, and the pressure increased by 4.3 mb. The transparency of the atmosphere is increased by~ 3%. This is due to the decline in solar activity or an accidental coincidence is hard to say, because considered a very small period of time.

Criteria of selected data also affect the results. Only days of optical observations of cosmic rays are selected – 15 days per month.

## 5. CONCLUSION

Comprehensive measurements of the atmosphere yielded a continuous series of data on temperature, pressure and atmospheric transparency that as the analysis can be used to determine the causes affecting the change in weather.

From the perspective of explanation of the data, we can talk about strong seasonal cyclonic activity in region of Yakutsk associated with recently more frequent removal of warm fronts from the southern region, i.e. redistribution of flow of warm air, which ultimately increases the average air temperature during winter. On the other hand, it can be associated with cycles of Solar activity [5, 6]. Table 2, shows the correlation of solar activity $R_i$ and intensity of galactic cosmic rays $I_{gcr}$ with geophysical and optical parameters of the atmosphere. According to the data of the table, there is a correlation between temperature and solar activity. Moreover, transparency of the atmosphere and number of cyclones do not depend on solar activity. Conversely, galactic cosmic rays indicate the GCR correlation with parameters such as atmospheric transmission coefficient and the number of cyclones, but the randomness of such a coincidence is quite large.